\documentclass[prd,twocolumn,showpacs,preprintnumbers,amsmath,amssymb]{revtex4}

\usepackage{graphicx}
\usepackage{dcolumn}
\usepackage{bm}

\def\be{\begin{eqnarray}}
\def\ee{\end{eqnarray}}
\def\bea{\begin{eqnarray}}
\def\eea{\end{eqnarray}}

\def\bT{{\bf b}_\perp}
\def\bs{{\bf b}_\perp^2}

\def\DT{{\bf \Delta}_\perp}
\def\Ds{{\bf \Delta}_\perp^2}
\def\0T{{\bf 0}_\perp}
\begin{document}


\title{Some Inequalities for the Generalized Parton Distribution
$E(x,0,t)$}

\author{Matthias Burkardt}
 \affiliation{Department of Physics, New Mexico State University,
Las Cruces, NM 88003-0001, U.S.A.}

\date{\today}

\begin{abstract}
We discuss some constraints on the $x$ and $t$-dependence of 
$E(x,0,t)$ that arise from positivity bounds in the impact parameter
representation. In addition, we show that $E(x,0,0)$ for the nucleon 
vanishes for $x\rightarrow 1$ at least as rapidly as $(1-x)^4$.
Finally we provide an inequality that limits the contribution
from $E$ to the angular momentum sum rule.
\end{abstract}

\maketitle
\section{Introduction}
Generalized parton distributions (GPDs) \cite{m,ji,r} are hybrid
quantities that have features in common both with form factors and 
with the usual parton distribution functions (PDFs). They are 
defined as non-forward ($p\neq p^\prime$) matrix elements of the
same operator $\hat{O} \equiv \int \frac{dx^-}{2\pi}e^{ix^-\bar{p}^+x}
\bar{q}\left(-\frac{x^-}{2}\right)
\gamma^+ q\left(\frac{x^-}{2}\right)$ whose forward matrix elements
(i.e. expectation value) yield the usual parton distributions
\be\label{eq:defHE}
\left\langle p^\prime \left|
\hat{O}
\right|p\right\rangle
&=&H(x,\xi,\Delta^2)\bar{u}(p^\prime)\gamma^+ u(p)
\\
&+& E(x,\xi,\Delta^2)\bar{u}(p^\prime)
\frac{i\sigma^{+\nu}\Delta_\nu}{2M} u(p)
.\nonumber
\ee
Over the last few years there has been a strong interest in 
GPDs and meanwhile many observables have been identified that can
be linked to them (for a recent review see Ref. \cite{d}).
One of the most interesting observables that can be linked 
to GPDs is a quantity that has identified been identified
with the total (spin plus orbital) angular momentum carried by the 
quarks in the nucleon \cite{ji}
\be \label{j}
\langle J_q\rangle = \frac{1}{2} \int_0^1 dx\, x \left[H_q(x,0,0)+
E_q(x,0,0)\right],
\ee
where the subscript $q$ indicates that Eq. (\ref{j}) holds for
each quark flavor separately. Of course, $H_q(x,0,0)=q(x)$ is
well known for the relevant values of $x$, but little is known
about $E(x,0,0)$.

Although GPDs can be probed in Compton scattering experiments,
they usually enter experimentally measurable cross sections only in 
terms of some integrals and therefore there may be some difficulties
in unambiguously extracting GPDs from Compton scattering data.
It is therefore desirable to use as many model-independent theoretical
constraints as possible to help pin down the data on GPDs.

One class of such constraints are positivity constraints, where one
uses the fact that any state in a Hilbert space has a non-negative 
norm. By using carefully constructed states one can thus derive
inequalities relating physical observables.

\section{Positivity Constraints in Impact Parameter Space}

In the case of GPDs, the impact parameter space representation
\cite{r} turns out to be very useful, since GPDs (for $\xi=0$)
become diagonal in that basis \cite{s,me:1st,jr,diehl,ijmpa}.
Parton distributions in impact parameter space are related to 
GPDs via a simple Fourier transform
(throughout this paper, we use a notation where parton
distributions in impact parameter space are denoted by script letters)
\bea
{\cal H}(x,\bT ) &=&
\int \!\!\frac{d^2 \DT}{(2\pi)^2}
H(x,0,-\Ds)e^{i\bT\cdot\DT}\\
\tilde{\cal H}(x,\bT ) &=&
\int \!\!\frac{d^2 \DT}{(2\pi)^2}
\tilde{H}(x,0,-\Ds)e^{i\bT\cdot\DT}\nonumber\\
{\cal E}(x,\bT ) &=&
\int \!\!\frac{d^2 \DT}{(2\pi)^2}
E(x,0,-\Ds)e^{i\bT\cdot\DT}.\nonumber
\eea
In Ref. \cite{aussie}, it was observed that the probabilistic
interpretation of parton distributions in impact parameter space
implies the positivity bound 
\be
\frac{1}{2M}\left| {\bf \nabla}_{\bT} {\cal E}(x,\bT) \right|
\leq {\cal H}(x,\bT).
\label{ineq1}
\ee
Here $x>0$; for $x<0$ a similar inequality with ${\cal H}\rightarrow
-{\cal H}$ holds.
In Ref. \cite{poby}, an even stronger bound
\be
\frac{1}{(2M)^2}\left| {\bf \nabla}_{\bT} {\cal E}(x,\bT) \right|^2
\leq \left|{\cal H}(x,\bT)\right|^2
-\left|\tilde{\cal H}(x,\bT)\right|^2
\label{ineq2}
\ee
was derived. 
Although Eqs. (\ref{ineq1}) and (\ref{ineq2}) are rigorous, their
practical use has been rather limited so far since they are  
relations between several unknown quantities. In this paper, 
we will manipulate these positivity bounds into a form, where they 
should be more directly applicable to phenomenology. For this 
purpose we first take Eq. (\ref{ineq2}) and simply integrate over
impact parameter. Since the inequality is preserved under this 
operation, and since the norm is invariant under Fourier 
transformation (e.g. $\int d^2\bT \left|{\cal H}(x,\bT)\right|^2
= \int \frac{d^2\DT}{(2\pi)^2}\left| H(x,0,-\Ds)\right|^2$), one
immediately finds
\be \label{ineqf}
& &\!\!\!\!\! \!\!\!\!\! \!\!\!\!\! \!\!\!
\int_{-\infty}^0 \!\! \!\!\!dt \frac{|t|}{(2M)^2}\left|{E}(x,0,t) 
\right|^2\\
& &\quad\quad\quad\leq\int_{-\infty}^0  \!\!\!\!\!dt\left\{ \left|{H}(x,0,t)\right|^2
-\left|\tilde{H}(x,0,t)\right|^2\right\}.
\nonumber
\ee
Similar expressions can be derived by repeating this procedure with
additional powers of $\left|\bT\right|$ in the integrand.
While this result immediately deals with the GPDs rather than
parton distributions in impact parameter space, its usefulness is
still limited by the fact that it involves 3 unknown functions and
therefore leaves too much room for model dependence.
It would be much more useful if we had constraints relating $E(x,0,t)$
to some known functions, such as the forward PDFs $q(x)$ and
$\Delta q(x)$. Deriving such relations will be the main goal in the 
rest of this paper.

For this purpose, we first introduce impact parameter
dependent parton distributions for quarks with spins parallel
[${\cal H}_+(x,\bT)$] and anti-parallel [${\cal H}_-(x,\bT)$]
to the nucleon spin (longitudinally polarized target)
\bea
{\cal H}_\pm(x,\bT) &\equiv& \frac{1}{2}\left[
{\cal H}(x,\bT) \pm\tilde{\cal H}(x,\bT)\right].
\eea
In terms of ${\cal H}_\pm $, Eq. (\ref{ineq2}) can be expressed
in the form
\be
\frac{1}{2M}\left| {\bf \nabla}_{\bT} {\cal E}(x,\bT) \right|
\leq 2\sqrt{{\cal H}_+(x,\bT) {\cal H}_-(x,\bT)}.
\label{ineq3}
\ee
Integrating the l.h.s. of Eq. (\ref{ineq3}) over the 
transverse plane yields [rotational invariance implies
${\cal E}(x,\bT) ={\cal E}(x,b) $]
\be
&\frac{1}{2M}&\!\!\!\! \int \!\!d^2\bT \left| {\bf \nabla}_{\bT} 
{\cal E}(x,\bT) \right| = \frac{\pi}{M}
\int_0^\infty db b \left| \partial_b {\cal E}(x,b)\right|
\nonumber\\
&&\geq \frac{\pi}{M}\left|\int_0^\infty db {\cal E}(x,b)\right|
= \frac{1}{2M} \left|
\int d^2\bT \frac{{\cal E}(x,b)}{\left|\bT\right|}\right|
\nonumber\\
&&= \frac{1}{4\pi M} \left|\int d^2 \DT \frac{E(x,0,-\Ds)}
{\left|\DT\right|}\right| \nonumber\\
&&= \frac{1}{4M}\left|
\int_{-\infty}^0 dt \frac{E(x,0,t)}{\sqrt{-t}}\right|,
\label{ineq4}\ee
where we used $\int d^2\bT \frac{e^{-i\bT\cdot\DT}}{
\left|\bT\right|} = \frac{2\pi}{\left|\DT\right|}$.

When integrating the r.h.s. of Eq. (\ref{ineq3}), we use 
the Schwarz inequality to obtain
\be\label{ineq5}
& &\!\!2\int d^2\bT \sqrt{{\cal H}_+(x,\bT) {\cal H}_-(x,\bT)}
\\
& &\leq 2 \sqrt{\left(\int d^2\bT {\cal H}_+(x,\bT)\right)
\left(\int d^2\bT^\prime {\cal H}_-(x,\bT^\prime)\right)}
\nonumber\\
& &= 2\sqrt{ q_+(x)q_-(x)} ,\nonumber
\ee
where $q_\pm(x) \equiv \frac{1}{2}\left( q(x)\pm \Delta q(x)\right)$
are the parton distribution for quarks with spin
parallel (anti-parallel) to the nucleon spin. Combining Eqs.
(\ref{ineq4}) and (\ref{ineq5}) yields
\be
\frac{1}{8M}\left|\int_{-\infty}^0 dt \frac{E(x,0,t)}{\sqrt{-t}} 
\right| < 
\sqrt{ q_+(x)q_-(x)},\label{ineq6}
\ee
which is one of the results of this paper. 
Like Eqs. (\ref{ineq1}) and (\ref{ineq2}), this result
holds for each quark flavor.

While Eq. (\ref{ineq6}) is weaker than our starting
point (\ref{ineq2}), it may still be of more use at this point
because it contains only one unknown quantity [$E(x,0,t)$]
and relates it to $q_\pm(x)$, which are much better known from
parton phenomenology

Although the r.h.s. of Eq. (\ref{ineq6}) involves only
known quantities, the l.h.s. still involves an integral.
For practical applications it may be more useful to have
an inequality that contains the unintegrated GPD $E$.
For this purpose we now multiply Eq. (\ref{ineq3}) by
$\left|\bT\right|$ and integrate. For the l.h.s. we find
\be
&\frac{1}{2M}&\!\!\!\! \int \!\!d^2\bT \left|\bT\right|
\left| {\bf \nabla}_{\bT} 
{\cal E}(x,\bT) \right| = \frac{\pi}{M}
\int_0^\infty db b^2 \left| \partial_b {\cal E}(x,b)\right|
\nonumber\\
&&\geq \frac{2\pi}{M}\left|\int_0^\infty db b{\cal E}(x,b)\right|
= \frac{1}{M} \left| \int d^2\bT {\cal E}(x,b)\right|
\nonumber\\
&&= \frac{1}{M} \left|E(x,0,0)\right|,
\label{ineq7}\ee
which involves $E(x,0,0)$, i.e. the quantity entering the angular
momentum sum rule \cite{ji}, directly.

On the r.h.s. one can invoke the Schwarz inequality in
different ways, and we choose to apply it in the form
\be\label{ineq8}
& &\!\!2\int d^2\bT \sqrt{\bs{\cal H}_+(x,\bT) {\cal H}_-(x,\bT)}
\\
& &\leq 2 \sqrt{\left(\int d^2\bT \bs {\cal H}_+(x,\bT)\right)
\left(\int d^2\bT^\prime {\cal H}_-(x,\bT^\prime)\right)}
\nonumber\\
& &= 2\sqrt{ \left. 4\frac{d}{dt}H_+(x,0,t)\right|_{t=0}q_-(x)} ,
\nonumber
\ee
where $H_\pm\equiv \frac{1}{2}\left(H\pm \tilde{H}\right)$.
Combining Eqs. (\ref{ineq3}),(\ref{ineq7}), and (\ref{ineq8})
we thus obtain
\be
\frac{1}{4M} \left|E(x,0,0)\right| 
\leq \sqrt{ q_-(x)\left. \frac{d}{dt}H_+(x,0,t)\right|_{t=0}} .
\label{ineq9}
\ee
While we do not know the slope of $H_+(x,0,t)$, we know some
general features. In particular, one expects that 
the transverse width of GPDs vanishes as $x\rightarrow 1$ \cite{ijmpa}
\be
\frac{\left. \frac{d}{dt}H_+(x,0,t)\right|_{t=0}}
{H_+(x,0,0)} \sim (1-x)^2 \quad \mbox{for}\quad x\rightarrow 1.
\label{width}
\ee
The reason for the vanishing of the transverse width for 
$x\rightarrow 1$
is that the variable conjugate to $\DT$ is the impact 
parameter $\bT$, which is measured w.r.t. the $\perp$ center
of momentum. The latter is related to 
the distance from the active quark  to the center of momentum of the
spectators (which we denote by ${\bf B_\perp}$ via the relation
$\bT = (1-x) {\bf B_\perp}$. The distance 
$\left|{\bf B_\perp}\right|$ between the active quark and the
spectators should be roughly equal to the size of the
nucleon or less. Being rescaled by a factor $(1-x)$,
the typical scale for $\bT$ is therefore only $
(1-x)$ times that size, which leads to Eq. (\ref{width}).

Making use of Eq. (\ref{width}) in Eq, (\ref{ineq9}) thus yields
\be
\left. \frac{d}{dt}H_+(x,0,t)\right|_{t=0}\sim (1-x)^{2+n_+}
 \quad \mbox{for}\quad x\rightarrow 1
\ee
where $n_\pm$ characterizes the behavior of $q_\pm(x)$ for 
$x\rightarrow 1$
\be
q_\pm(x) \sim (1-x)^{n_\pm}  \quad \mbox{for}\quad x\rightarrow 1 .
\ee
For example, if $n_+=3$ and $n_-=5$ (based on hadron helicity 
conservation \cite{BBS}), then
\be
E(x,0,0) \sim (1-x)^{1+\frac{n_++n_-}{2}}=(1-x)^5
 \quad \mbox{}\quad 
\ee
for $x\rightarrow 1$.
Even if there is a small contribution to the negative helicity
distribution $q_-(x)$ that vanishes with the same power as the
positive helicity distribution $q_+(x)$, i.e. if $n_+=n_-=3$,
then $E(x,0,0)$ would still
behave like $(1-x)^4$ and therefore vanish
faster than $H(x,0,0)$ as $x\rightarrow 1$. In either case we find
\be
\lim_{x\rightarrow 1} \frac{E(x,0,0)}{H(x,0,0)} =0.
\ee

For applications to the angular momentum sum rule (\ref{j}),
we can also try to convert Eq. (\ref{ineq9}) into a statement about
the $2^{nd}$ moment of $E$. Upon multiplying Eq. (\ref{ineq9}) by
$|x|$ and integrating from $-1$ to $1$ 
(antiquarks correspond to $x<0$), one finds
\be \label{ineq10}
&&\!\!\frac{1}{4M}\left|\int dx E(x,0,0) x\right|  \\ &&
\leq\frac{1}{4M}
\int dx \left|E(x,0,0)\right| |x| \nonumber\\ &&\leq
\int dx \sqrt{ xq_-(x)\left. \frac{d}{dt}xH_+(x,0,t)\right|_{t=0}}
\nonumber\\
&& \sqrt{\left(\int dx\,xq_-(x)\right)\left(\int dx^\prime
\left. \frac{d}{dt}x^\prime H_+(x^\prime ,0,t)\right|_{t=0}\right)},
\nonumber\ee
which contains only one unknown on the r.h.s., namely the slope
of the second moment of $H_+$. 
To illustrate that this inequality may provide some useful bounds,
let us insert some rough figures: not distinguishing between 
different flavors (i.e. implicitly adding all quark flavors)
we approximate: 
$\int dx xq_-(x)\approx \int dx xq_+(x) \approx \frac{1}{4}$ and
$\int dx^\prime \left. 
\frac{d}{dt}x^\prime H_+(x^\prime ,0,t)\right|_{t=0}
= \int dx  xq_+(x)\frac{R_+^2}{6}\approx \frac{1}{4}\frac{R_+^2}{6}$, 
where $R_+^2$ is the rms-radius corresponding to $\int dx x H_+$. 
We do not know the value of $R_+$ but it should be on the order of 
the rms 
radius of the nucleon. In fact, $R_+$ should be smaller than that 
since the slope of the form factor should decrease for increasing
$x$-moments (see the discussion following Eq. (\ref{width}) and
also Ref. \cite{n})
, i.e. we approximate $R_+\approx 0.5 fm$. Inserting these 
rough figures, we find
$\left|\int dx E(x,0,0)x\right| \leq \frac{R_+M}{\sqrt{6}}\approx 1$.
Although this is not a very strong constraint, a better estimate
may be available once the slope of the second moment of $H$ is
known for different quark flavors.

\section{Summary}
We started from positivity constraints for parton distributions
in impact parameter space (\ref{ineq2}) and derived several new
positivity constraints on GPDs (\ref{ineqf}),(\ref{ineq6}),
(\ref{ineq9}), and (\ref{ineq10}).
Although the new constraints are weaker than the staring inequality
(\ref{ineq2}), the new inequalities may be more useful since 
they can be applied directly in momentum space, where the data is
obtained. One of the new inequalities (\ref{ineq6}) relates 
$\int \frac{dt}{\sqrt{|t|}} E(x,0,t)$ directly to the (forward) 
parton distributions $q(x)$ and $\Delta q(x)$ and therefore provides 
a direct constraint on the shape of $E$. 

The third inequality that we derived (\ref{ineq9}) is a bound on
$E(x,0,0)$. Unfortunately, it involves the slope of $H(x,0,t)$ for
$t=0$, which is currently not known.

If hadron helicity conservation (HHC) holds and $q_-(x)\sim (1-x)^5$
for quarks with spin antiparallel to the nucleon spin 
then $E(x,0,0)\sim (1-x)^5$ as $x\rightarrow 1$ for valence
quarks in the nucleon. Even if HHC is violated, and $q_-(x)$ has 
a small component that vanishes  only like $(1-x)^3$ 
then $E$ still vanishes faster than the leading valence
distributions, i.e. $E(x,0,0)\sim (1-x)^4$.
This is a consequence of the fact that the positivity
constraints in impact parameter representation (\ref{ineq1}) and 
(\ref{ineq2})
involve ${\bf \nabla_{\bT}}$ together with the fact
that the $\perp$ width
of GPDs goes to zero as $x\rightarrow 1$.
Knowing that $E(x,0,0)$ vanishes faster than $q(x)$ near be useful
in estimating the contribution from $E$ to the angular momentum 
sum rule.

Finally we derived an inequality that can be used to constrain
the second moment of $E(x,0,0)$. The only unknown in this
inequality is the rms-radius for the second moment of $H$. 

{\bf Acknowledgements:}
This work was supported by the DOE under grant number 
DE-FG03-95ER40965.

\bibliography{ineq.bbl}
\end{document}